\documentclass[twocolumn]{aastex631}

\usepackage{amsmath}

\begin{document}

\title{A transdimensional sampling framework for pulsar timing noise modelling}

\author[0000-0003-3432-0494]{Valentina Di Marco}
\affiliation{School of Physics, University of Melbourne, Parkville, VIC 3010, Australia}
\affiliation{OzGrav: The ARC Center of Excellence for Gravitational Wave Discovery, Australia}

\author[0000-0002-0287-2270]{Nir Guttman}
\affiliation{School of Physics and Astronomy, Monash University, Vic 3800, Australia}
\affiliation{OzGrav: The ARC Center of Excellence for Gravitational Wave Discovery, Australia}

\author[0000-0002-5455-3474]{Matthew T. Miles}
\affiliation{Department of Physics and Astronomy, Vanderbilt University, 2301 Vanderbilt Place, Nashville, TN 37235, USA}
\affiliation{OzGrav: The ARC Center of Excellence for Gravitational Wave Discovery, Australia}

\author[0000-0002-9583-2947]{Andrew Zic}
\affiliation{CSIRO, Space and Astronomy, PO Box 76, Epping, NSW 1710, Australia}
\affiliation{OzGrav: The ARC Center of Excellence for Gravitational Wave Discovery, Australia}

\author[0000-0002-7285-6348]{Ryan M. Shannon}
\affiliation{Centre for Astrophysics and Supercomputing, Swinburne University of Technology, Hawthorn VIC 3122, Australia}
\affiliation{OzGrav: The ARC Center of Excellence for Gravitational Wave Discovery, Australia}

\author[0000-0002-4418-3895]{Eric Thrane}
\affiliation{School of Physics and Astronomy, Monash University, Vic 3800, Australia}
\affiliation{OzGrav: The ARC Center of Excellence for Gravitational Wave Discovery, Australia}

\begin{abstract}

A careful characterisation of the noise processes in pulsar timing data is a prerequisite for pulsar timing array experiments. 
While single-pulsar noise analyses are crucial for both gravitational-wave searches and astrophysical studies, they are often computationally intensive and rely on running and comparing multiple fixed noise models.
We present \texttt{tPTABilby}, a transdimensional Bayesian inference framework for single-pulsar noise analysis built on the \texttt{Bilby} library. 
The method flexibly models a wide range of noise processes like radiometer noise, pulse-phase jitter, intrinsic red noise, dispersion measure variations, and chromatic interstellar medium effects.
By employing transdimensional sampling, \texttt{tPTABilby} simultaneously infers the number and type of active noise sources, providing a unified treatment of model selection and parameter estimation. 
We validate the methodology through simulations with known injected noise models, demonstrating accurate recovery of model probabilities and calibrated posterior distributions. 
We then apply this approach to a single pulsar, PSR J1713$+$0747, from a MeerKAT Pulsar Timing Array (MPTA) dataset, analysing the data with both \texttt{tPTABilby} and \texttt{Enterprise}, and subsequently compare the results with existing MPTA analyses through posterior predictive checks of the inferred noise spectra. 
Our results highlight the flexibility of transdimensional approaches to single-pulsar noise analysis, demonstrating consistency with standard fixed-model methods while providing a more statistically robust framework, and present \texttt{tPTABilby} as a simple and reproducible approach for PTA inference.

\end{abstract}


\section{Introduction} \label{sec:intro}

Precision timing of millisecond pulsars provides a powerful tool: to probe astrophysical processes including extreme gravity and the interstellar medium \citep{Shao_2023, Geyer_2017}, to test theories of gravity \citep{Kramer_2006, Freire_2024}, and to search for a nanohertz gravitational-wave background \citep{Jenet_2005, Hobbs_2010}. 
A prerequisite for these analyses is an accurate characterisation of the noise processes that contribute to the pulsar timing residuals. 
These processes include radiometer noise, pulse-phase jitter, intrinsic spin noise, dispersion measure variations, chromatic interstellar medium effects, and solar wind \citep{Cordes_2010, Wang_2015, Goncharov_2020}. 
Single-pulsar noise analysis, in which the noise properties of individual pulsars are modelled independently, therefore plays a central role in pulsar timing array (PTA) experiments \citep{Reardon_2023_null_hyp, Agazie_2023_noise, EPTA_pulsar_noise, Miles_2024_noise}. 
Robust single-pulsar noise analysis is essential not only for optimising sensitivity to a common gravitational-wave background but also for improving our understanding of the various astrophysical noise processes that affect pulsar timing data.

Single-pulsar noise analyses model pulsar timing residuals obtained after fitting a deterministic timing model, and then explore a high-dimensional parameter space to characterise stochastic noise properties and compare different noise models within a Bayesian framework.
They are currently performed using
\texttt{Enterprise}\footnote{https://gitlab.in2p3.fr/epta/enterprise/-/tree/master/enterprise} \citep{enterprise} and \texttt{TempoNest}\footnote{https://github.com/aparthas3112/Temponest} \citep{Lentati_2013}. 
\texttt{Enterprise} is a general Bayesian pulsar timing analysis code in which one defines a generative model for the timing residuals by specifying deterministic timing parameters and stochastic processes such as white noise and red noise.
The code constructs likelihoods from these signal components and samples the joint posterior over all parameters using external samplers, allowing both noise and timing parameters to be inferred simultaneously.
\texttt{TempoNest} is a Bayesian analysis package built on \texttt{tempo2}\footnote{https://bitbucket.org/psrsoft/tempo2}\citep{Hobbs_2006, Edwards_2006} and nested sampling (utilising MultiNest \citep{Feroz_2008}) that jointly fits the non-linear timing model and stochastic noise processes, and enables model selection between competing noise descriptions through evaluation of the Bayesian evidence.

However, a limitation of this approach is that the appropriate complexity of the noise model is not known a priori. In practice, single-pulsar noise analyses therefore require analysts to define and compare a large number of competing noise models, before selecting a model deemed sufficient to describe the data. This manual model-selection procedure can be computationally expensive and may introduce implicit biases, particularly when many similar models are compared or when model choices are informed by preliminary inspection of the data. Recent work has highlighted how workflows that prune models using posterior results from the same data can introduce bias and advocated alternatives in which model complexity is inferred within a single analysis, for example through spike-and-slab–type model-averaging constructions \citep{Van_Haasteren_2024_B}.

In this paper, we introduce \texttt{tPTABilby}, a transdimensional Bayesian inference method for single-pulsar noise analysis.
The framework is designed to be modular and interoperable with existing PTA software, and we envisage future extensions interfacing with emerging tools such as \texttt{Discovery}\footnote{\url{https://github.com/nanograv/discovery}}, facilitating broader integration within PTA analysis pipelines.
\texttt{tPTABilby} is built on \texttt{tBilby} \citep{Tong_2025}, a transdimensional extension of the \texttt{Bilby} \citep{Ashton_2019} inference library.
It is designed to provide a streamlined and extensible inference scheme for PTA noise modelling.
The package builds on the existing ecosystem of PTA inference tools, such as \texttt{Enterprise}, and is designed to complement existing tools while streamlining noise analyses. 
Its primary design goals are flexibility and ease of use. 
New noise sources can be incorporated with minimal effort, as the code automatically builds a complex model structure that enumerates all possible model combinations using the same functions as \texttt{Enterprise}. 
This design makes it straightforward to extend the approach with additional noise processes.

The \texttt{tPTABilby} code is built around a transdimensional sampling framework that allows the number and type of noise sources to be inferred from the data.
In conventional pulsar-timing noise analyses, the number of free parameters is fixed a priori, and model comparison requires explicit evidence evaluation across separate analyses. In contrast, \texttt{tPTABilby} treats the number of sources $N$ as an additional parameter to be inferred. When the number of sources is free, the dimensionality of the parameter space varies with $N$, and the posterior must capture both the parameter values and their count. 
This approach provides a natural and efficient way to perform simultaneous parameter estimation and model selection, while automatically incorporating Occam’s razor through Bayesian evidence, favouring simpler models unless the data justify additional complexity.
This will become even more important with large and sensitive PTA experiments, such as those planned for the DSA and SKA \cite[][]{Shannon_2025}. 

In PTAs, transdimensional samplers have been used to model non-stationary and non-power-law noise by varying the number of wavelets or spectral control points in response to the data \citep{Ellis_2016}. 
Again, because transdimensional samplers intrinsically penalise overly complex models through the Occam factor, they allow the data to dictate the level of model complexity required.

Transdimensional techniques have also been implemented within \texttt{PTMCMC} \citep{PTMCMC} through a discrete model index parameter (denoted \texttt{nmodel}), which enables reversible-jump moves between a finite set of predefined models. 
Our approach is conceptually related in that it performs joint model selection and parameter estimation within 
a single inference run. 
In principle, one could recover an equivalent 
model-marginalised result within \texttt{PTMCMC} by explicitly defining all $2^K$ candidate noise-model combinations and allowing reversible-jump moves between them.
However, rather than switching between a manually specified list of alternative models, we treat each candidate noise process as having its own on/off switch. 
This construction automatically explores the full combinatorial model space within a unified nested-sampling framework, reducing manual model management and simplifying bookkeeping.

In Section~\ref{sec:method}, we describe the methodology, including the noise models and transdimensional framework. 
In Section~\ref{sec:sims}, we assess the validity of the method through controlled simulations. 
In Section~\ref{sec:real_data}, we apply the method to real data from the MPTA-DR3 dataset, focusing on PSR J1713+0747 and restricted to observations prior to the 2021 profile event and present posterior predictive checks of the inferred noise spectra. 

Finally, in Section~\ref{sec:conclusion}, we discuss our findings and outline directions for future development.

\section{Method} \label{sec:method}

In PTA analyses, pulsar timing noise arises from multiple stochastic processes, and the specific combination of these processes differs from pulsar to pulsar.
In some single-pulsar analyses, only a small subset of noise terms such as achromatic red noise and dispersion-measure variations are included in the model \citep[e.g.,][]{Arzoumanian_2020}, while other studies incorporate some additional components such as chromatic noise, system-dependent white noise terms, or band/group noise \citep[e.g.,][]{Antoniadis_2022, Reardon_2023_gw}.

Traditional Bayesian approaches therefore require each candidate noise combination to be specified and evaluated separately, which becomes computationally challenging as the number of possible models grows combinatorially.

To address this, we implement a transdimensional inference framework in which the inclusion of each candidate noise sources is governed by a binary indicator variable. 
Instead of fixing the model structure in advance, we allow the sampler to explore across all possible model combinations, dynamically switching noise sources on or off during the run.
Each configuration corresponds to a different effective dimensionality of the parameter space, since activating a noise source introduces its associated parameters, while deactivating it removes them.  

Formally, the transdimensional model can be expressed as
\begin{equation}
M(\boldsymbol{\theta}, \{n_i\}) = \sum_{i=1}^{K} n_i\, f_i(\boldsymbol{\theta}_i),
\quad n_i \in \{0,1\}.
\end{equation}
Here, the model $M$ denotes the total stochastic noise power spectral density (PSD) which defines the stochastic process in the frequency domain; the likelihood, however, depends on the corresponding time-domain covariance matrix $C$ constructed from $M$ \citep{Lentati_2013, Van_Haasteren_2014, enterprise}.
$M$ has units of $\mathrm{s}^2\,\mathrm{Hz}^{-1}$. 
Each function $f_i(\boldsymbol{\theta}_i)$ represents the PSD contribution of the $i$-th noise process, parameterised by $\boldsymbol{\theta}_i$, and the binary indicator $n_i \in \{0,1\}$ specifies whether that process is included in the model.

For a given configuration $\{n_i\}$, the model defines the total noise covariance matrix, which we write explicitly as
\begin{equation}
    C(\boldsymbol{\theta},\{n_i\})=C_{\mathrm{EF}}(\boldsymbol{\theta}_{\mathrm{EF}})+\sum_{i=1}^{K} n_i\, C_i(\boldsymbol{\theta}_i),
\end{equation}
where $C_{\mathrm{EF}}$ is the contribution of the EFAC white-noise term, which is always included in the model (i.e. it is not subject to the transdimensional switch).
The EFAC term accounts for potential miscalibration of the time-of-arrival (TOA) measurement uncertainties. In practice, each TOA has an associated uncertainty $\sigma_{\mathrm{TOA}}$, and EFAC rescales these uncertainties by a multiplicative factor, such that the white-noise variance becomes $\mathrm{EFAC}^2 \sigma_{\mathrm{TOA}}^2$. This corresponds to a diagonal contribution to the covariance matrix,
\begin{equation}
C_{\mathrm{EF}} = \mathrm{diag}\left( \mathrm{EFAC}^2 \sigma_{\mathrm{TOA}}^2 \right).
\end{equation}
We choose to always include EFAC in the model because imperfect calibration of measurement uncertainties is expected at some level for all pulsars, and it provides a minimal and necessary degree of flexibility in modelling white noise.

Meanwhile, $C_i(\boldsymbol{\theta}_i)$ is the covariance contribution associated with the $i$-th additional stochastic noise process.
Each covariance contribution $C_i(\boldsymbol{\theta}_i)$ is constructed from the corresponding power spectral density $f_i(\boldsymbol{\theta}_i)$ via the standard Fourier-domain representation of stationary stochastic processes. In particular, using the Fourier design matrix $F$, the covariance can be written as
\begin{equation}
C_i(\boldsymbol{\theta}_i) = F \, \Phi_i(\boldsymbol{\theta}_i) \, F^\mathrm{T},
\end{equation}
where $\Phi_i$ is a diagonal matrix.
Each diagonal element of $\Phi_i$ is proportional to $f_i(\boldsymbol{\theta}_i)$ evaluated at the corresponding discrete Fourier frequency. 
This construction ensures that the total covariance matrix is the linear superposition of the individual stochastic processes.

The likelihood of the timing residuals is then given by the standard PTA Gaussian likelihood,
\begin{equation}
\begin{split}
\mathcal{L}(d \mid \boldsymbol{\theta}, \{n_i\}) ={}&
\frac{1}{\sqrt{(2\pi)^{N_d}\,\det C(\boldsymbol{\theta}, \{n_i\})}} \\
&\times \exp\!\left[
-\tfrac{1}{2}\, r^{\top} C^{-1}(\boldsymbol{\theta}, \{n_i\}) r
\right].
\end{split}
\end{equation}
where $r$ denotes the timing residuals and $N_{d}$ is the number of data points, i.e., the number of times of arrival (TOAs).

Priors are specified at two levels. The binary switches are given a prior
\begin{equation}
    \pi(\{n_i\}) = \prod_{i=1}^K \pi(n_i), \quad \pi(n_i=1) = \pi(n_i=0) = \tfrac{1}{2},
\end{equation}
unless prior knowledge favours particular noise sources. The continuous parameters for each noise sources follow their own astrophysically motivated priors, so that
\begin{equation}
    \pi(\theta \mid \{n_i\}) = \prod_{i=1}^K \pi(\theta_i)^{n_i}.
\end{equation}
As explained in \cite{Tong_2025}, parameters associated with inactive noise sources ($n_i=0$) are sampled as ``ghost parameters'': they are not included in the likelihood evaluation but ensure dimensional consistency across the parameter space. 
After marginalising over these ghost parameters, the posterior is equivalent to a brute-force analysis over all fixed models, but achieved within a single unified inference run, with the conceptual and statistical advantages outlined in the Introduction.

The joint posterior distribution over the noise parameters $\theta$ and the binary indicators $\{n_i\}$ is
\begin{equation}
    p(\theta, \{n_i\} \mid d) \propto \mathcal{L}(d \mid \theta, \{n_i\}) \, \pi(\theta \mid \{n_i\}) \, \pi(\{n_i\}),
    \end{equation}
where $d$ denotes the pulsar timing residuals, $\mathcal{L}$ is the likelihood, $\pi(\theta \mid \{n_i\})$ are the priors on the continuous noise parameters, and $\pi(\{n_i\})$ is the prior over model configurations.

Finally, the marginal posterior over the noise parameters is obtained by summing over all model configurations,
\begin{equation}
    p(\theta \mid d) = \sum_{\{n_i\}} p(\theta, \{n_i\} \mid d),
\end{equation}
so that the analysis simultaneously gives parameter estimates and model probabilities. This construction removes the need to run and combine multiple fixed-model analyses in post-processing, thereby simplifying bookkeeping and reducing the risk of inconsistent workflows. The sampler naturally spends more time in models with stronger evidence, while overly complex models are penalised through Bayesian Occam’s razor. 
Since \texttt{tPTABilby} is based on the \texttt{tBilby} library, it retains the modular flexibility of \texttt{bilby} but adds support for variable-dimensional model spaces, offering compatibility with multiple samplers, including nested samplers such as \texttt{dynesty} and parallel-tempered MCMC samplers such as \texttt{ptemcee}.

\subsection{Method implementation}

The method is organised into three main components:

\begin{enumerate}
    \item \textbf{Model construction.}  
    A dedicated model registry maintains and constructs the set of candidate noise processes considered in the analysis.
    Each noise process is implemented as an \texttt{enterprise} signal block with its own parameterisation and priors. 
    A model-definition class specifies which noise sources are optional, and all admissible combinations of active and inactive noise sources are enumerated. 
    Each unique configuration is assigned a binary model key, which maps directly to a corresponding \texttt{enterprise} PTA object.

    \item \textbf{Likelihood evaluation.}  
    A custom likelihood class interfaces the PTA models with the sampler. 
    The likelihood is defined globally, but at each evaluation step the binary indicator parameters select a specific model configuration. 
    The corresponding PTA object is retrieved using the model key, the relevant subset of parameters is extracted, while parameters associated with inactive noise sources (which we refer to as ghost parameters) are ignored, and the log-likelihood is evaluated using the \texttt{enterprise} likelihood implementation. 
    Priors for the \texttt{Enterprise} PTA parameters are converted into a \texttt{Bilby}-compatible prior dictionary using the \texttt{get\_bilby\_prior\_dict} utility provided by the \texttt{enterprise\_warp} package.\footnote{\url{https://github.com/bvgoncharov/enterprise_warp}}
    This design enables transdimensional exploration of the model space within a single likelihood function. 
    The binary indicator parameters are treated as discrete sampling dimensions with associated priors, enabling joint exploration of continuous noise parameters and model dimensionality.

    \item \textbf{Posterior analysis.}  
    Posterior samples are grouped according to their model keys, allowing direct estimation of posterior model probabilities from their sampling frequencies. Parameter posteriors can be analysed either conditionally on a given model or marginalised over all models to produce model-averaged posterior distributions.
\end{enumerate}
While we focus here on conditional posteriors for comparison with existing analyses, the sampler output naturally represents the fully model-averaged posterior
\begin{equation}
p(\boldsymbol{\theta} \mid d) = \sum_m p(\boldsymbol{\theta} \mid d, m)\, p(m \mid d),
\end{equation}
which can be obtained by marginalising over the sampled noise-model configurations.

\subsection{Sampling and output}

Inference is performed using nested sampling (\texttt{dynesty}), which simultaneously outputs posterior samples for both the noise parameters and the binary indicators, as well as Bayesian evidence information that allows posterior model probabilities to be estimated.
Posterior samples naturally concentrate on the most probable configurations as disfavored models receive negligible sampling weight, leading to highly efficient exploration of model space. 
Posterior samples provide continuous parameter estimates, while the frequencies of different binary configurations reflect model probabilities. 
In this way, robust model selection and parameter estimation are achieved within a single, unified inference run.

\section{Model validation} \label{sec:sims}

\subsection{Simulations}

The main stochastic noise processes we expect to be present in pulsar timing are broadly categorised into white noise, arising mainly from radiometer noise and pulse jitter \citep{Wang_2015, Kulkarni_2024}, and red noise, which dominates at low frequencies and includes both achromatic noise sources such as spin noise \citep{Shannon_2010, Cordes_2010, Lasky_2015} and chromatic contributions such as dispersion measure variations \citep{Keith_2013} and interstellar scattering \citep{Rickett_1990, Cordes_2010,Kulkarni_2025}. 
Additional sources include solar wind-induced fluctuations in electron column density \citep{You_2007,Tiburzi_2016,Madison_2019}, band and group noise linked to propagation or instrumental effects \citep{Cordes_2016, van_Straten_2013, Lentati_2016}, system-dependent jumps from instrumentation changes \citep{Verbiest_2016, Kerr_2020, Reardon_2023_null_hyp}, and transient phenomena such as chromatic dips and extreme scattering events \citep{Goncharov_2020}. 
All red-noise processes are specified in terms of a (one-sided) power spectral density (PSD) for the timing residuals,
\begin{equation}
S_r(f)=\frac{A^2}{12\pi^2}
\left(\frac{f}{1\,\mathrm{yr}^{-1}}\right)^{-\gamma}\,\mathrm{yr}^3,
\end{equation}
where $A$ is the amplitude referenced to $1\,\mathrm{yr}^{-1}$ and $\gamma$ is the spectral index.
In the likelihood, this continuous PSD is discretised onto a Fourier basis with bin widths $\Delta f_k$, yielding per-mode variances
\begin{equation}
\rho_k = S_r(f_k)\,\Delta f_k,
\end{equation}
so that for the standard uniformly spaced Fourier frequencies one recovers $\Delta f_k = 1/T$, with $T$ the observation timespan.

For the purpose of our tests and validations, in our simulations we consider only the following noise sources: The three parameters for white noise (EFAC, EQUAD and ECORR) as well as the timing noise (TN), the dispersion measure (DM) variations and the chromatic noise (CH).
In all analyses presented in this work, the chromatic noise (CH) is implemented with a fixed chromatic index of $-4$, i.e., the process scales with observing frequency as $\nu^{-4}$.

We simulate 300 datasets of the pulsar J0437$-$4715 using the \texttt{libstempo}\footnote{\url{https://github.com/vallis/libstempo}} Python interface to \texttt{Tempo2} \citep{Hobbs_2006}.
This pulsar is one of the most precisely timed and extensively studied millisecond pulsars in pulsar timing array experiments, making it a representative test case for validating noise inference methods. 
The simulations adopt observing characteristics representative of typical pulsar timing array datasets, such as those from the Parkes Pulsar Timing Array (PPTA), including realistic observing cadences, time spans, and measurement uncertainties.
Starting from a set of real \texttt{Tempo2}-format pulsar ephemerides, we generate idealised times-of-arrival using the \texttt{make\_ideal} function and apply an EFAC factor to the TOA uncertainties. 
We then inject random combinations of EQUAD, ECORR, intrinsic timing noise (TN), dispersion measure variations, and chromatic noise (CH), with each noise source included with equal probability. 
For each of the 300 simulations, noise parameters are drawn from uniform distributions with ranges listed in Table~\ref{tab:sim_params}.
The same uniform prior distributions are adopted in the subsequent analysis.

\begin{deluxetable}{lc}
    \tablecaption{Parameter ranges used for noise injection in the pulsar simulations. Amplitudes are given in $\log_{10}$ seconds.\label{tab:sim_params}}
    \tablehead{
    \colhead{Parameter} & \colhead{Range (uniform)}
    }
    \startdata
        EFAC & $0.5$--$1.5$ \\
        $\log_{10}\mathrm{ECORR}$ & $-6.5$\;\;$-6$ \\
        $\log_{10}\mathrm{EQUAD}$ & $-6.5$\;\;$-6$ \\
        $\log_{10}A_{\mathrm{TN}}$ & $-14.2$\;\;$-13.8$ \\
        $\gamma_{\mathrm{TN}}$ & $1.5$\;\;$3$ \\
        $\log_{10}A_{\mathrm{DM}}$ & $-13.6$\;\;$-13.2$ \\
        $\gamma_{\mathrm{DM}}$ & $3.5$\;\;$4.5$ \\
        $\log_{10}A_{\mathrm{CH}}$ & $-14.8$\;\;$-14.4$ \\
        $\gamma_{\mathrm{CH}}$ & $3.5$\;\;$4.5$ \\
    \enddata
\end{deluxetable}

\subsection{Validation}

We validate the transdimensional inference method using the 300 simulated datasets with known injected noise models and parameters. 
The validation focusses on two complementary aspects: (i) the statistical consistency of posterior model probabilities for noise-model selection, and (ii) the statistical consistency  of posterior distributions for the noise parameters. 

\subsubsection{Model-selection statistical consistency} \label{sec:application}

To assess whether the transdimensional framework correctly recovers the injected noise models, we examine the statistical consistency of posterior model probabilities across the ensemble of simulations.
For each simulation, we run \texttt{tPTABilby} allowing the number and type of active noise sources to vary. 
Posterior samples are grouped according to their model configuration, and the posterior probability assigned to a given model is estimated as the fraction of samples associated with that configuration.

For each simulation, the injected noise model is identified by converting the known active noise sources into the same binary representation used in the \texttt{tPTABilby} inference (see Appendix A). In practice, we assign a number to each model. 
The posterior probability assigned to the injected model,
\begin{equation}
    P_{\mathrm{injected}} = \frac{\# \text{ posterior samples in injected model}}{\# \text{ total posterior samples}},
\end{equation}
provides a per-simulation measure of how strongly the sampler supports the true model.

For the validation, simulations are grouped into bins according to the value of $P_{\mathrm{injected}}$. 
For each bin, we compute the empirical fraction of simulations whose data were generated from the corresponding injected model. 
If the posterior model probabilities are well calibrated, this empirical frequency should match the assigned posterior probability. 
This relationship can be visualised by a validation curve, which should follow the diagonal line $y = x$.

\begin{figure}
    \centering
    \includegraphics[width=0.9\linewidth]{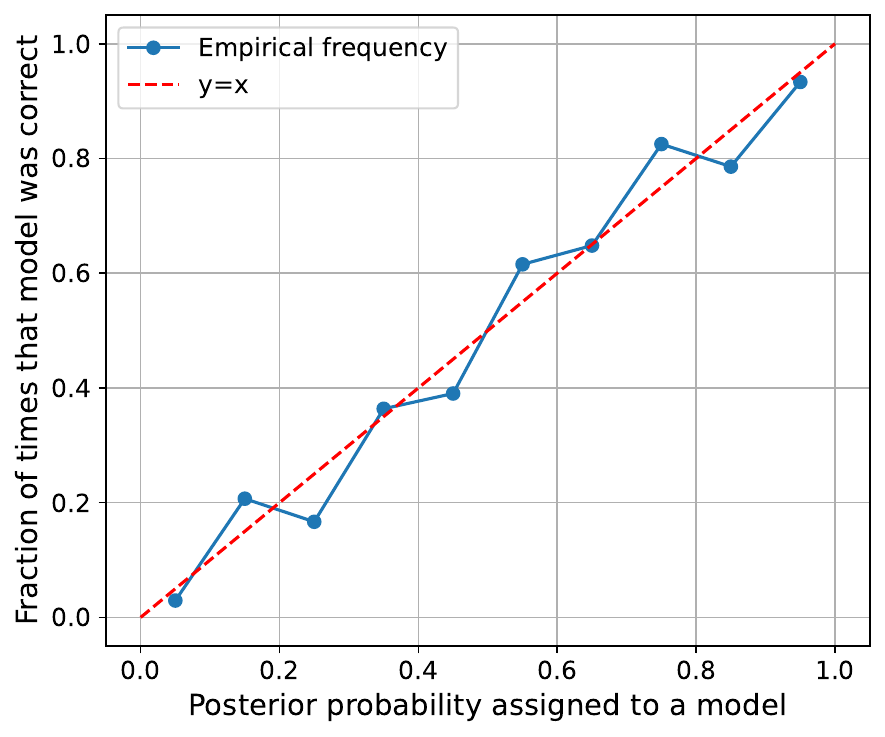}
    \caption{Validation curve for posterior model probabilities. 
    The posterior probability assigned to the injected noise model is binned across simulations, and the empirical frequency with which the injected model is correct is computed for each bin. 
    If posterior model probabilities are well behaved, the curve should follow the diagonal line $y=x$ (dashed red line).}

    \label{fig:calibration-plot}
\end{figure}

Deviations from the diagonal would indicate over- or under-confidence in model selection and may point to issues in evidence estimation, prior weighting, or transdimensional sampling behaviour. 
The resulting validation curve is shown in Figure~\ref{fig:calibration-plot}, demonstrating that the posterior model probabilities are well behaved across the explored model space with no significant deviations observed.

\subsubsection{Parameter posterior validation}

In addition to model-selection performance, we test the statistical consistency of posterior distributions for the noise parameters using a posterior quantile (or rank) method.
For each simulation, the true parameter values used to generate the data are compared against the corresponding posterior samples obtained from the inference.

For each injected parameter, we compute the empirical cumulative distribution function of its posterior and evaluate it at the injected value.
This yields a posterior quantile, or p-value, defined as the fraction of posterior samples with values less than the injected parameter.
If the inference procedure is well behaved, the injected values should be uniformly distributed within the posterior, and the resulting $p$-values should follow a uniform distribution on the interval $[0,1]$.

\begin{figure*}
\centering

    \includegraphics[width=0.32\textwidth]{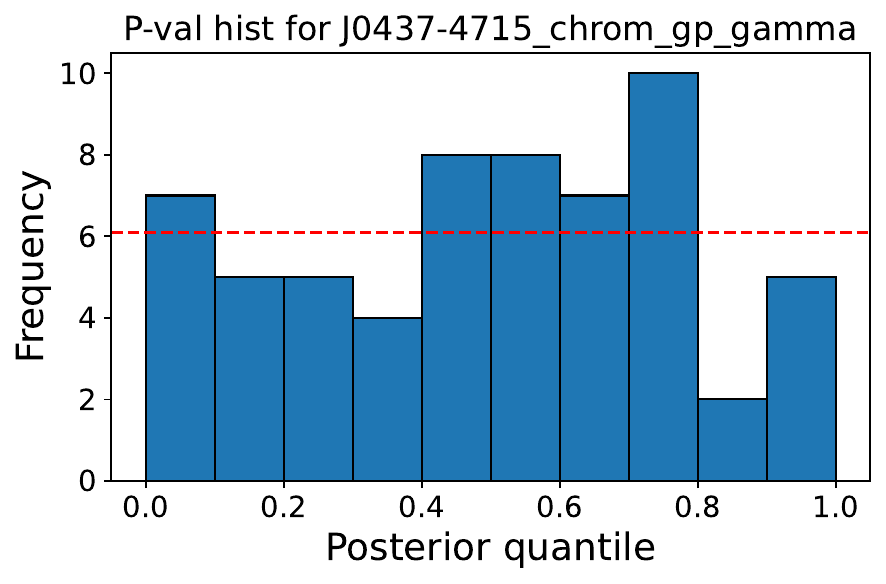}
    \includegraphics[width=0.32\textwidth]{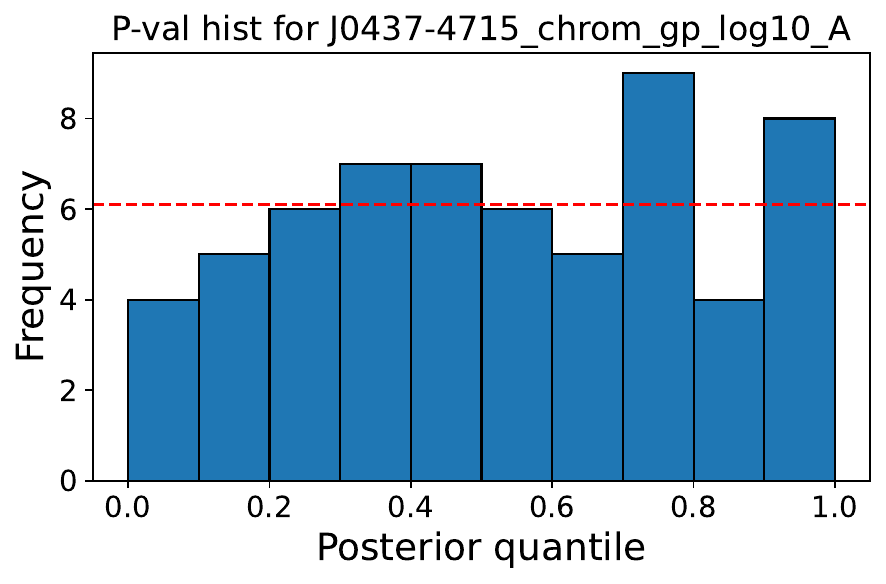}
    \includegraphics[width=0.32\textwidth]{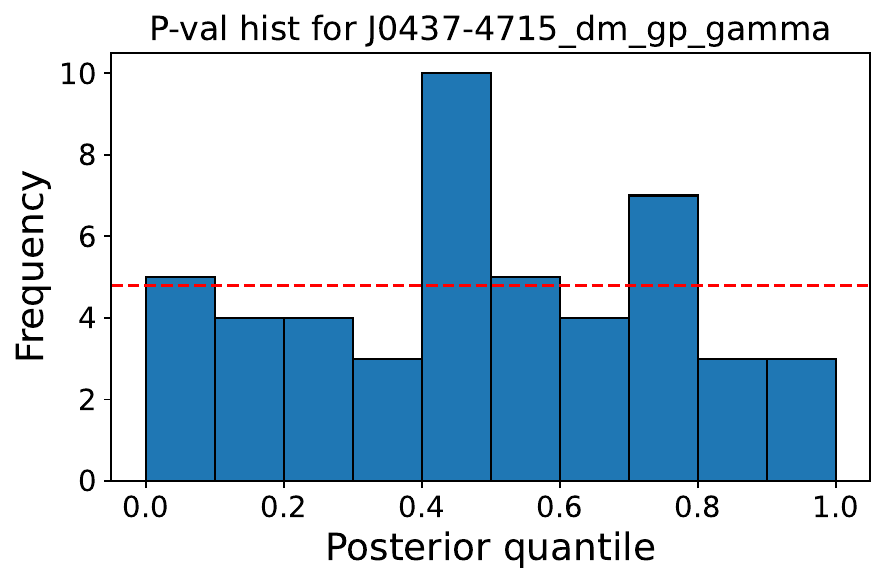}
    
    \vspace{0.5em}
    
    \includegraphics[width=0.32\textwidth]{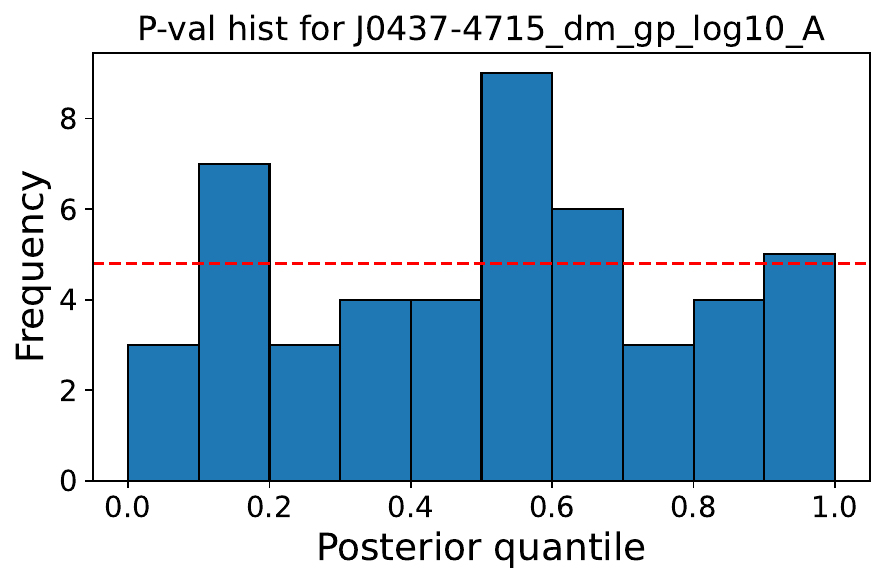}
    \includegraphics[width=0.32\textwidth]{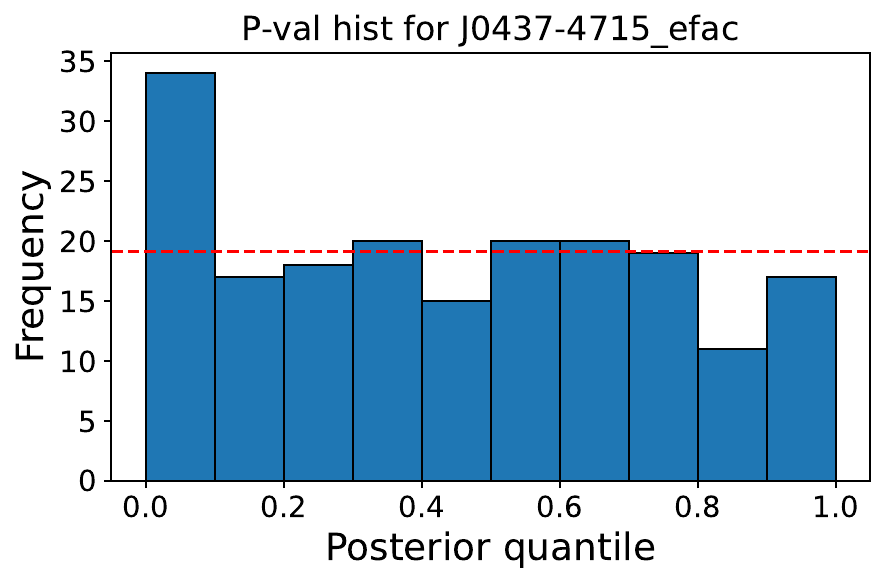}
    \includegraphics[width=0.32\textwidth]{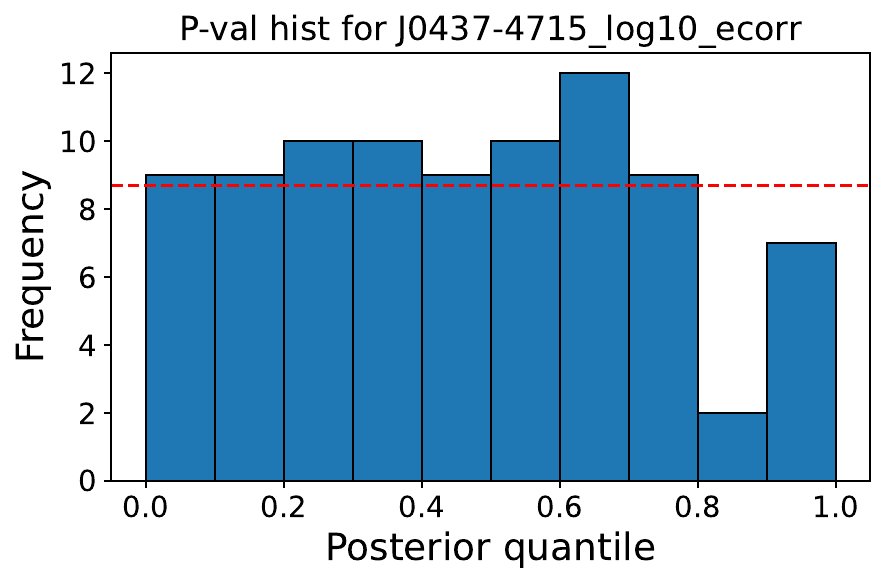}
    
    \vspace{0.5em}
    
    \includegraphics[width=0.32\textwidth]{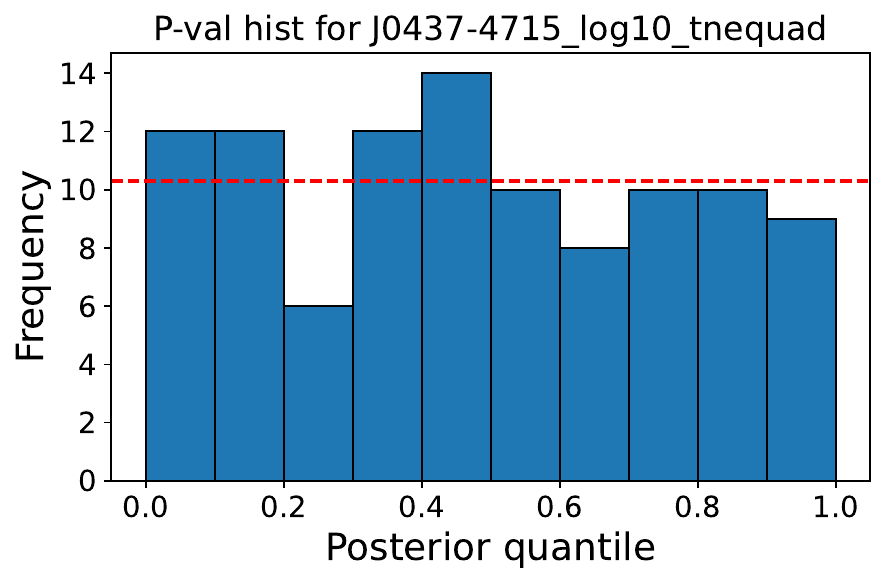}
    \includegraphics[width=0.32\textwidth]{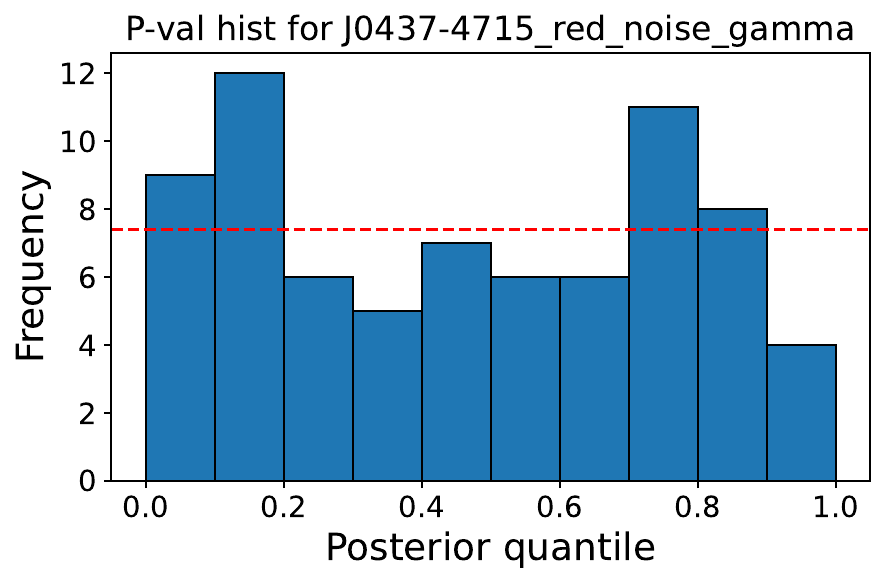}
    \includegraphics[width=0.32\textwidth]{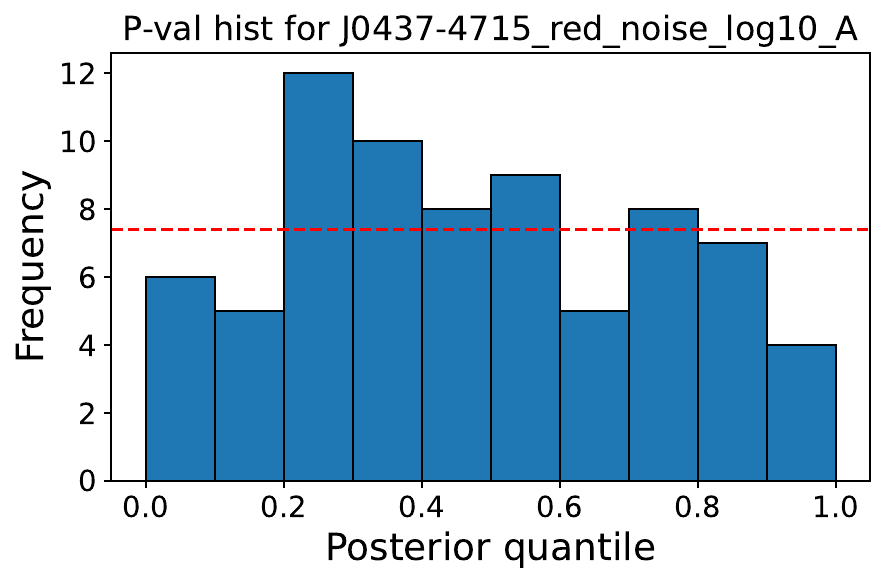}

    \caption{Posterior quantile ($p$-value) histograms for continuous noise parameters across all simulations. For each parameter, the posterior cumulative distribution function is evaluated at the injected value, yielding a posterior quantile. If the posterior distributions are well calibrated, these quantiles should be uniformly distributed on $[0,1]$. The dashed red line marks the expected uniform distribution.}
    \label{fig:histograms}
\end{figure*}

We apply this test to each noise parameter across all simulations, producing posterior quantile histograms and performing Kolmogorov--Smirnov tests to assess consistency with uniformity.
This validation probes the statistical consistency of the continuous parameter posteriors independently of the model-selection process, and is sensitive to potential biases arising from likelihood construction, prior specification, or the treatment of ghost parameters.
The results of this test are shown in Fig.~\ref{fig:histograms}.

\section{Application to MeerKAT Dataset}\label{sec:real_data}

To further validate \texttt{tPTABilby} on real data, we perform posterior
predictive checks of the pulsar noise spectrum using data obtained as part of the MeerKAT Pulsar Timing Array (MPTA) project.
We analyse an internal dataset representative of the forthcoming MPTA DR3 release (hereafter MPTA-DR3), focusing on a single pulsar and noting that it is preliminary and may differ from the final data set.  Using MPTA-DR3, we assess the consistency of \texttt{tPTABilby} on real observations through two complementary comparisons. Our analysis focuses on observations of PSR~J1713$+$0747. The data set was produced using the same methods used to produce MPTA DR2 \cite[][]{Miles_2024_noise}. The data set only includes observations prior to the 2021 profile event, in which the radio pulse profile had an abrupt shape change that introduced large systematic errors in the pulse arrival times \cite[][]{Singha_2021}.

We draw posterior samples of the noise parameters and construct the
corresponding timing-residual power spectral density (PSD), expressed
in units of $\mathrm{s}^2\,\mathrm{Hz}^{-1}$. Posterior draws are aggregated
across all visited noise-model configurations and weighted by their
posterior model probabilities, thereby obtaining a model-marginalised
posterior predictive distribution.

We visualise these spectra as ensembles of thin lines representing
individual draws, together with the median and 68\% credible interval
of the marginalised PSD. For comparison, we also overplot the median
spectrum predicted by the preferred single model, defined as the model
with the highest posterior probability. In the case of J1713$+$0747, we show that the preferred-model
median lies well within the marginalised credible band, indicating
that uncertainty in model selection does not qualitatively affect the
inferred noise spectrum.

First, we compare posterior predictive PSDs obtained with \texttt{tPTABilby} against those produced by an analysis of the same data using Enterprise, adopting the same noise model. 
This comparison isolates differences arising solely from the inference framework and sampling strategy.
Second, we compare \texttt{tPTABilby} results against the preferred noise model adopted by the MPTA collaboration, providing a consistency check against an independently validated analysis.

Importantly, all analyses---\texttt{tPTABilby}, \texttt{Enterprise}, and MPTA ---identify the same preferred noise model for J1713+0747, consisting of a chromatic noise process and a steep-spectrum red noise component (modelled with a fixed spectral index of 13/3, corresponding to a gravitational-wave-background-like process).

\begin{figure}
    \centering
        \includegraphics[width=0.9\linewidth]{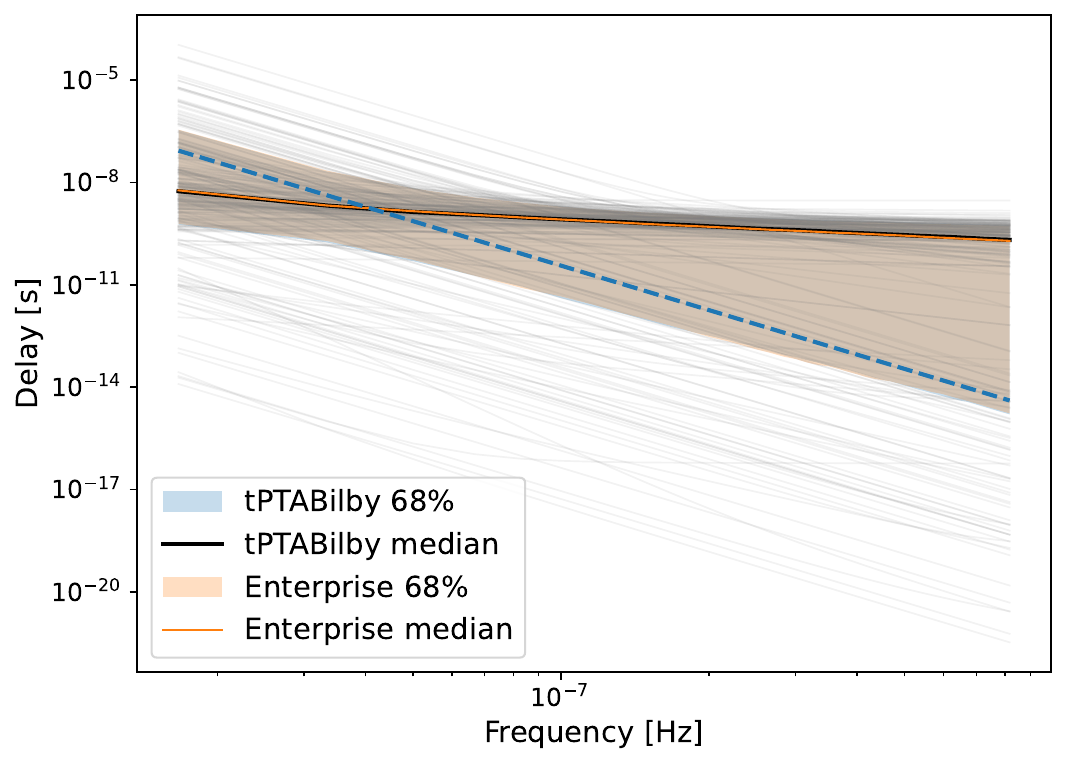}
        \caption{Posterior predictive power spectral densities for pulsar J1713$+$0747 inferred using \texttt{tPTABilby} and \texttt{Enterprise} under the same preferred noise model.
        Thin lines represent individual posterior predictive draws, while the solid line and shaded region indicate the median and 68\% credible interval, respectively. The dashed straight line represents a fiducial SGWB with amplitude $A = 10^{-14.35}$ and spectral index $\gamma = 13/3$.
        The close agreement between the two methods demonstrates consistent noise inference for a fixed model on real PTA data.
        }
        \label{fig:Enterprise_PPC}
\end{figure}

Figure~\ref{fig:Enterprise_PPC} shows the resulting posterior predictive PSDs, with shaded regions indicating the 68\% credible intervals and solid lines denoting the posterior medians.
The posterior predictive distributions obtained with \texttt{tPTABilby} and \texttt{Enterprise} are found to be in close agreement across the full frequency range, with overlapping credible regions and consistent median spectra.
This agreement demonstrates that \texttt{tPTABilby} reproduces consistent noise descriptions with \texttt{Enterprise} when applied to real PTA data.

\begin{figure}
    \centering
        \includegraphics[width=0.9\linewidth]{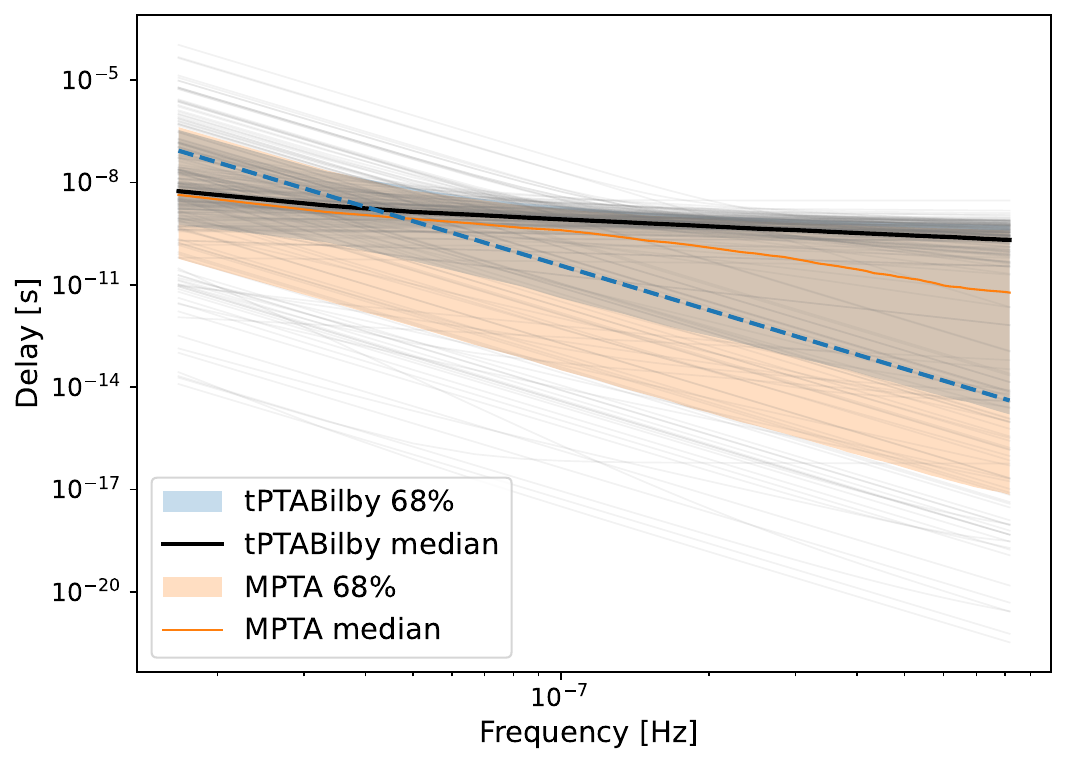}
        \caption{Posterior predictive power spectral densities for pulsar J1713$+$0747 inferred using \texttt{tPTABilby} and the MPTA collaboration results under the same preferred noise model.
        Thin lines represent individual posterior predictive draws, while the solid line and shaded region indicate the median and 68\% credible interval, respectively. The dashed straight line represents a fiducial SGWB with amplitude $A = 10^{-14.35}$ and spectral index $\gamma = 13/3$.
        The close agreement between the two methods demonstrates consistent noise inference for a fixed model on real PTA data.}
        \label{fig:Matt_PPC}
\end{figure}

In the second comparison, we compare the \texttt{tPTABilby} posterior predictive PSDs with those derived from the preferred noise models from the MPTA analysis.
Figure~\ref{fig:Matt_PPC} shows that the inferred PSDs are again broadly consistent, particularly at low frequencies where red-noise and stochastic gravitational-wave background contributions dominate.
Differences at higher frequencies are modest and fall within the respective posterior uncertainties, reflecting differences in the sampling methodology and software versions between the two analysis pipelines, rather than inconsistencies in inference.

Overall, the posterior predictive checks demonstrate that \texttt{tPTABilby} produces noise inferences consistent with both \texttt{Enterprise} and with existing MPTA collaboration results when applied to real data.
These checks and earlier simulations provide confidence that the transdimensional framework yields reliable posterior predictions while remaining compatible with established PTA analysis pipelines.

\begin{figure*}
\centering

    \includegraphics[width=0.32\textwidth]{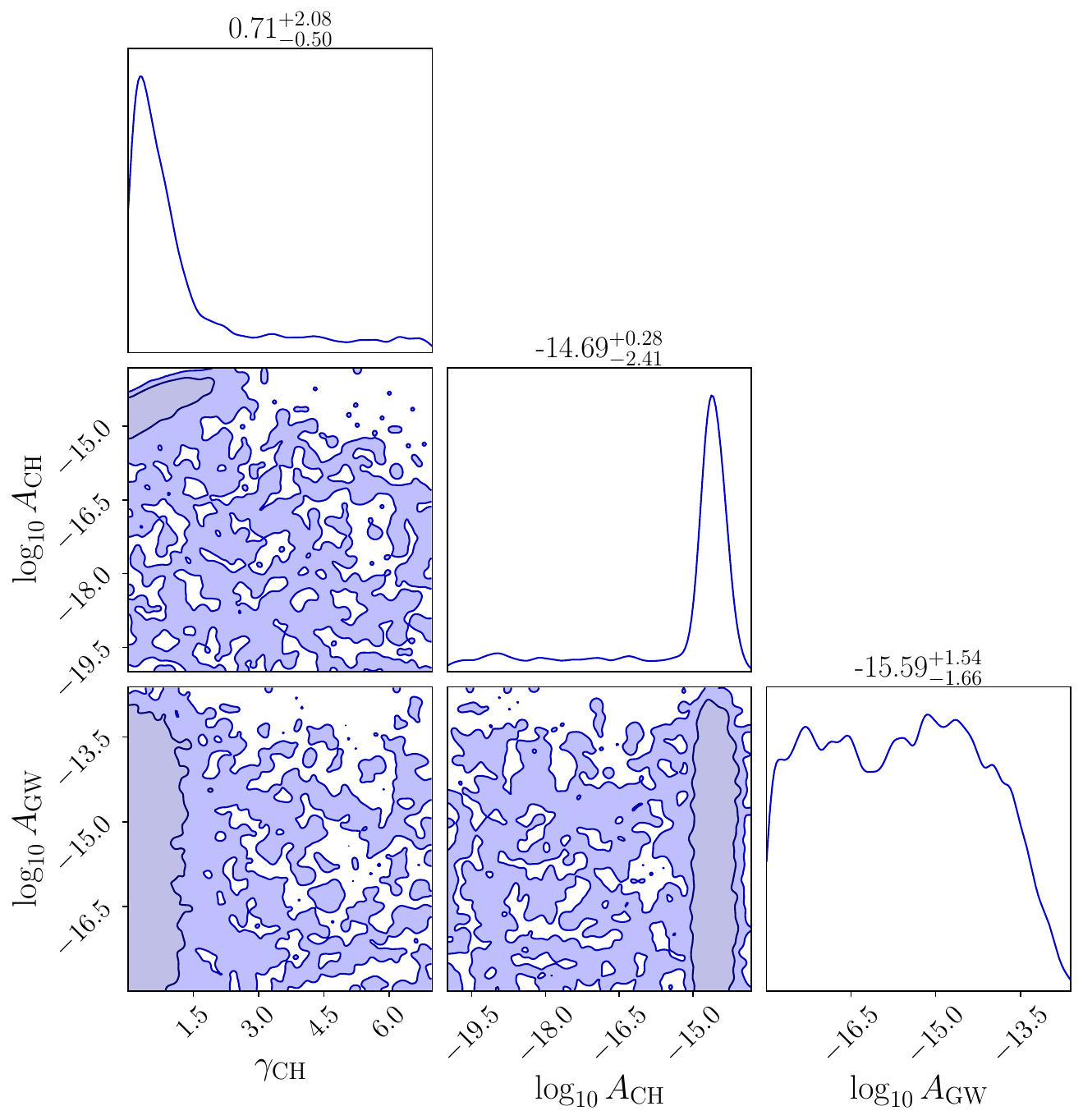}
    \includegraphics[width=0.32\textwidth]{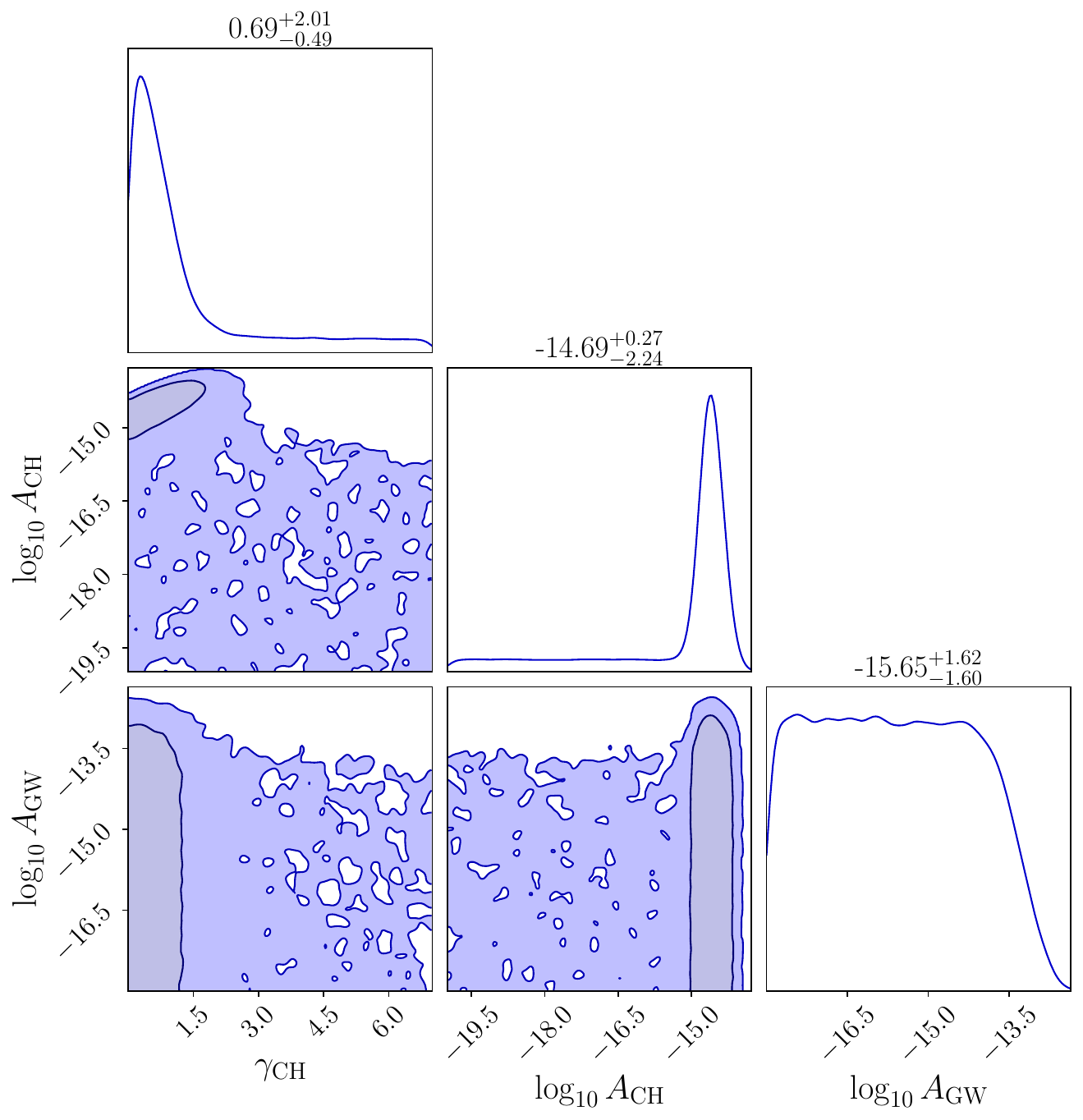}
    \includegraphics[width=0.32\textwidth]{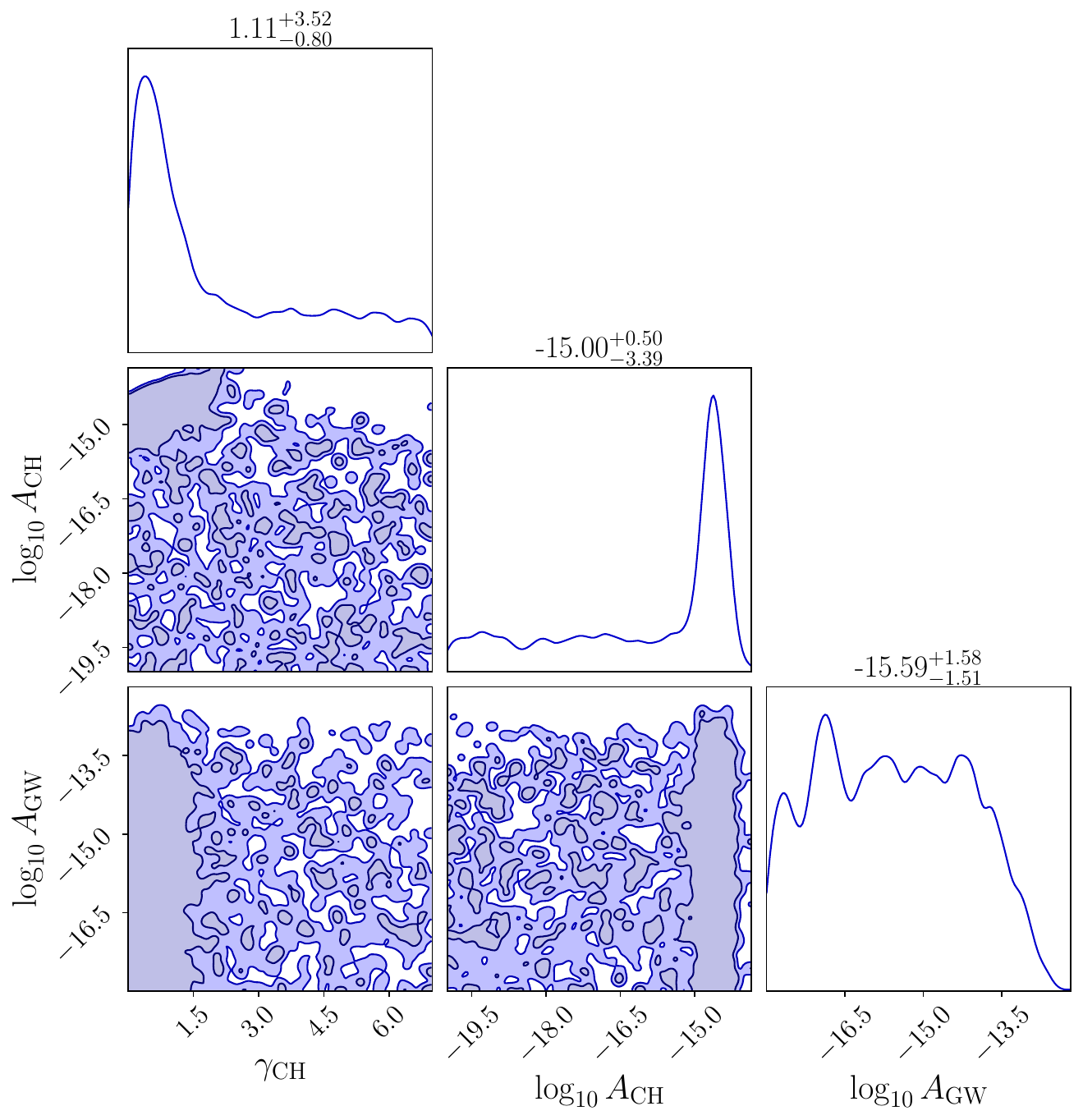}

    \caption{Marginal posterior distributions for selected noise parameters obtained with \texttt{tPTABilby} (left), \texttt{Enterprise} (centre), and the MPTA collaboration analysis (right), using the same noise model and data selection --- the model preferred by all analyses.}
\label{fig:Posteriors}
\end{figure*}

In Figure~\ref{fig:Posteriors} we also show marginal posterior distributions for the key noise parameters of the preferred model, obtained with \texttt{tPTABilby}, \texttt{Enterprise} and the MPTA data analysis.

\section{Discussion and conclusions} \label{sec:conclusion}

In this work we present \texttt{tPTABilby}, a transdimensional Bayesian inference framework for single-pulsar noise analysis in PTA datasets. 
The method is built on \texttt{tBilby} \citep{Tong_2025} and interfaces with \texttt{Enterprise} \citep{enterprise}, enabling a flexible specification of stochastic timing-noise sources. 
By introducing binary indicator variables for candidate noise processes, \texttt{tPTABilby} performs parameter estimation and model comparison within a single sampling run, producing posterior weights for competing noise-model configurations alongside posterior constraints on continuous noise parameters.

Several previous PTA noise analyses have relied on scanning over a discrete set of fixed noise models, selecting a single preferred model using Bayes factors, and then conditioning subsequent inference on that model. As discussed in \citet{Van_Haasteren_2024_B}, this procedure can lead to a form of circular or statistically inconsistent inference, since uncertainty in the noise model is not fully propagated into parameter estimation or detection statistics. In a formally Bayesian treatment, uncertainty over competing noise models should be marginalised rather than resolved by model selection.

In Bayesian terms, correct inference for parameters $\boldsymbol{\theta}$ given data $d$ requires marginalisation over the discrete set of noise models $m$,
\begin{equation}
p(\boldsymbol{\theta} \mid d) = \sum_m p(\boldsymbol{\theta} \mid d, m)\, p(m \mid d),
\end{equation}
whereas fixed-model approaches effectively approximate this as
\begin{equation}
p(\boldsymbol{\theta} \mid d) \simeq p(\boldsymbol{\theta} \mid d, m_{\mathrm{best}}),
\end{equation}
thereby neglecting uncertainty in the model structure. Conditioning on a single selected noise model can bias parameter estimates, overstate detection significance, and underestimate posterior uncertainties. While results obtained using this approach are likely to be qualitatively correct in many cases, they are statistically incomplete.

The transdimensional framework implemented here performs this marginalisation by construction, jointly sampling over noise-model configurations and their associated parameters within a single inference run. This ensures that noise-model uncertainty is consistently propagated into all posterior inferences and detection statistics. 

An important consequence of this more complete treatment is that marginalising over flexible noise models can reduce apparent detection significance, as probability mass is distributed over a broader set of competing explanations for the data.
In practice, this effect can be mitigated through the use of physically motivated prior probabilities on the inclusion of specific noise processes. Informative or hierarchical priors, potentially learned from population-level PTA analyses, can downweight implausible noise configurations while still allowing additional complexity when strongly supported by the data. Finally, while rigorous noise-model marginalisation is essential in the pre-detection regime, it is arguably even more critical in the post-detection era, where accurate characterisation of the gravitational-wave background amplitude, spectral index, and anisotropy depends sensitively on a robust and statistically consistent treatment of pulsar noise.

We validate the method using 300 simulated datasets based on J0437$-$4715 with known injected noise sources and parameters. 
In these tests, posterior probabilities assigned to injected models are well calibrated, and posterior quantile (rank) diagnostics for continuous parameters are consistent with the expected uniform behaviour. 
These results indicate that the transdimensional construction gives reliable posterior model probabilities and well-calibrated parameter posteriors for the PTA noise processes considered in this study.

We further assess the behaviour of \texttt{tPTABilby} on real data by applying it to the MEDR3 dataset. 
Posterior predictive checks of the inferred noise power spectral densities show good agreement between \texttt{tPTABilby} and an \texttt{Enterprise}-based analysis on the same data when adopting the same preferred noise model. 
In addition, comparisons with existing MPTA noise analysis  results show broadly consistent posterior predictive spectra and overlapping posterior constraints for key noise parameters under the preferred model. 
These results demonstrate that \texttt{tPTABilby} reproduces established single-pulsar noise inferences on real PTA data while providing a unified approach for exploring model uncertainty.

The main aim of \texttt{tPTABilby} is to provide a statistically correct, simple and reproducible analysis method for single-pulsar noise modelling, with an extensible design that facilitates the inclusion of additional noise processes.
Future work will extend the catalogue of supported stochastic noise sources and investigate applications to broader PTA analyses, including systematic studies across pulsar populations and extensions to models that incorporate more complex spectral shapes and non-stationary behaviour. 
Extensions to fully multi-pulsar analyses and to non-stationary or time-localised noise sources are also left to future work.
Finally, we also plan to explore adapting \texttt{tPTABilby} to interface with \texttt{Discovery}\footnote{\url{https://github.com/nanograv/discovery}}, enabling compatibility with emerging PTA software.

\section*{Acknowledgements}
We acknowledge and pay respects to the Elders and Traditional Owners of the land on which this work has been performed, the Bunurong, Wadawurrong and
Wurundjeri People of the Kulin Nation and the Wallumedegal People of the Darug Nation.
The MeerKAT telescope is operated by the South African Radio Astronomy Observatory, which is a facility of the National Research Foundation, an agency of the Department of Science and Innovation.
The authors are supported via the Australian Research Council (ARC) Centre of Excellence CE230100016.
E.T.\ and N.G\ are additionally supported through ARC Discovery Project DP230103088.

\section*{Data Availability}
This work uses data from the MeerKAT Pulsar Timing Array (MPTA) project. The dataset analysed is an internal, preliminary version representative of the forthcoming MPTA Data Release 3 (MPTA-DR3). The data are not yet publicly available but will be released as part of MPTA third data release.
The scripts used in this analysis are available on \url{https://github.com/valeaussie/tPTABilby.git}.

\appendix
\section{Noise-model catalogue and model indexing}\label{app:model_catalogue}

In \texttt{tPTABilby}, the stochastic-noise model space is defined by a set of $K$ candidate noise processes, each controlled by a binary indicator $n_i \in \{0,1\}$. The indicator vector $\mathbf{n} \equiv (n_1,\dots,n_K)$ specifies which processes are active in a given model configuration. For a fixed choice of candidate processes, the total number of possible models is
\begin{equation}
N_{\mathrm{models}} = 2^{K},
\end{equation}
where $K$ is the number of candidate noise processes included in the transdimensional analysis.
Table~\ref{tab:ni_mapping} provides the explicit correspondence between the binary indicators $n_i$ and the noise processes considered in this work.

\begin{table}
\centering
    \caption{Binary indicator mapping used to define the transdimensional noise-model space. A value $n_i=1$ activates the corresponding process; $n_i=0$ deactivates it.}
    \label{tab:ni_mapping}
    \begin{tabular}{lll}
    \hline
    Index $i$ & Indicator $n_i$ & Noise process \\
    \hline
    1 & $n_{\mathrm{EFAC}}$  & EFAC (TOA uncertainty scaling) \\
    2 & $n_{\mathrm{EQUAD}}$ & EQUAD (additional white noise) \\
    3 & $n_{\mathrm{ECORR}}$ & ECORR (epoch-correlated white noise) \\
    4 & $n_{\mathrm{TN}}$    & Timing noise (achromatic red noise) \\
    5 & $n_{\mathrm{DM}}$    & DM variations (chromatic $\nu^{-2}$ process) \\
    6 & $n_{\mathrm{CH}}$    & Chromatic red noise (generic chromatic process) \\
    \hline
\end{tabular}
\end{table}

\subsection{Binary encoding and model numbering}\label{app:model_numbering}

For convenience in analysis and visualisation, each binary configuration $\mathbf{n}$ is assigned a unique integer model index $m$. We adopt a binary-to-decimal mapping
\begin{equation}
m(\mathbf{n}) \equiv \sum_{i=1}^{K} n_i\,2^{\,i-1},
\qquad m \in \{0,1,\dots,2^{K}-1\}.
\label{eq:model_index}
\end{equation}
With this convention, $m=0$ corresponds to the model in which all candidate processes are inactive ($n_i=0$ for all $i$), while $m=2^{K}-1$ corresponds to the model in which all processes are active ($n_i=1$ for all $i$).
The index ordering is fully determined by the ordering of processes in Table~\ref{tab:ni_mapping}.
(If a subset of processes is used in a given analysis, $K$ in Eq.~\ref{eq:model_index} refers to that subset, and the total number of models is $2^{K}$.)

\subsection{\texttt{tPTABilby} outputs: posterior over models}\label{app:outputs}

A primary discrete output of \texttt{tPTABilby} is the posterior probability assigned to each model configuration. Posterior samples are grouped by their model index $m$, and the posterior model probability is estimated by the sampling frequency,
\begin{equation}
p(m \mid d) \approx \frac{N_m}{N_{\mathrm{tot}}},
\end{equation}
where $N_m$ is the number of posterior samples associated with model $m$ and $N_{\mathrm{tot}}$ is the total number of posterior samples.

In practice, we visualise $p(m \mid d)$ as a bar plot over model index $m$, where the horizontal axis enumerates the $2^{K}$ possible models (with $K$ the number of candidate noise processes used in the run) and the vertical axis shows the posterior sample count. 
This provides an immediate diagnostic of which noise-model configurations dominate the posterior and how strongly the data favour a small subset of the full model space.

\begin{figure}
    \centering
    \includegraphics[width=0.6\linewidth]{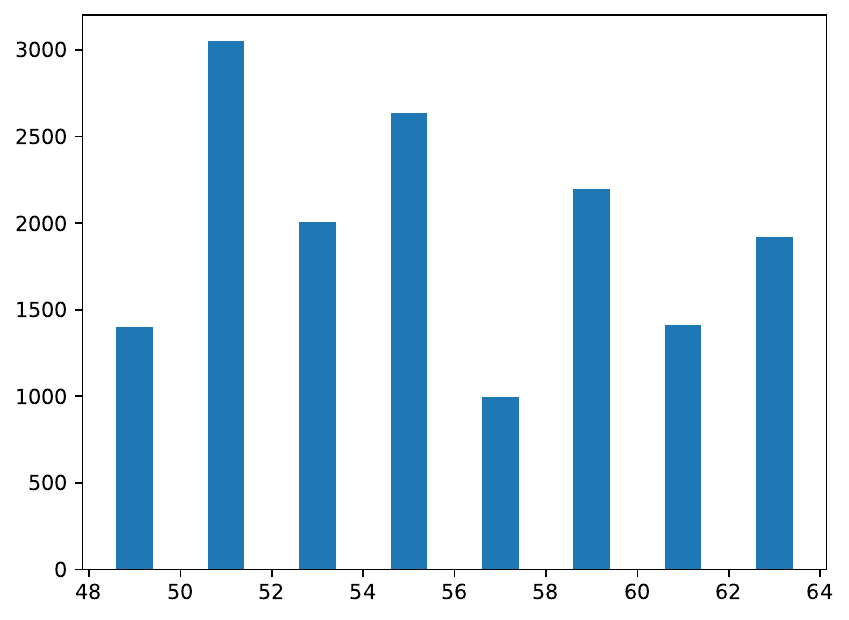}
    \caption{Example \texttt{tPTABilby} model-posterior summary. Each bar corresponds to a model index $m$ defined by Eq.~\ref{eq:model_index}, and the bar height shows the posterior mass (or equivalently the posterior sample count) assigned to that model. The total number of possible models is $2^{K}$, where $K$ is the number of candidate noise processes included in the transdimensional analysis.}
    \label{fig:model_posterior_barplot}
\end{figure}

\bibliography{refs}{}
\bibliographystyle{aasjournal}

\end{document}